\documentclass[aps,prb,twocolumn,showpacs,groupedaddress]{revtex4}
\usepackage[dvips]{epsfig}

\begin{document}
\title{Weak ferromagnetism with very large canting in a chiral lattice:
(pyrimidine)$_{2}$FeCl$_{2}$}

\author{R. Feyerherm, A. Loose}
\affiliation{Hahn-Meitner-Institut and Berlin Neutron Scattering
Center, 14109 Berlin, Germany}

\email{Feyerherm@hmi.de}

\author{T. Ishida, T. Nogami}
\affiliation{Department of Applied Physics and Chemistry, The
University of Electro-Communications, Chofu, Tokyo 182-8585,
Japan}

\author{J. Kreitlow, D. Baabe, F. J. Litterst, S. S\"ullow, H.-H. Klauss}
\affiliation{Institut f\"ur Metallphysik und Nukleare
Festk\"orperphysik, TU Braunschweig, 38106 Braunschweig, Germany}

\author{K. Doll}
\affiliation{Institut f\"ur Mathematische Physik, TU Braunschweig,
38106 Braunschweig, Germany}

\date{\today}

\begin{abstract}
The transition metal coordination compound
(pyrimidine)$_{2}$FeCl$_{2}$ crystallizes in a chiral lattice,
space group $I4_{1}22$ (or $I4_{3}22$). Combined magnetization,
M\"oss\-bauer spectroscopy and powder neutron diffraction studies
reveal that it is a canted antiferromagnet below $T_{N} = 6.4$ K
with an unusually large canting of the magnetic moments of
14$^\circ$ from their general antiferromagnetic alignment, one of
the largest reported to date. This results in weak ferromagnetism
with a ferromagnetic component of $1~\mu_B$. The large canting is
due to the interplay between the antiferromagnetic exchange
interaction and the local single-ion anisotropy in the chiral
lattice. The magnetically ordered structure of
(pyrimidine)$_{2}$FeCl$_{2}$, however, is not chiral. The
implications of these findings for the search of molecule based
materials exhibiting chiral magnetic ordering is discussed.

\end{abstract}



\maketitle

\section{Introduction}

One major research effort in the very active field of molecule
based magnetic materials \cite{ICMM2000,ICMM2002,Kahn,Miller,Gatt}
is directed towards synthesizing multifunctional compounds that
combine magnetism with a second physical property, such as
conductivity \cite{NatureCoro} or optical activity.\cite{Optic} In
this framework, extensive efforts are undertaken to find chiral
magnetic materials that exhibit magnetochiral dichroism (MChD), a
phenomenon first observed by Rikken and Raupach in a chiral
paramagnetic material. \cite{Rikken} Large MChD is expected in
materials that combine chirality and magnetic order. To date,
however, there is no evidence for large MChD, although a number of
chiral magnetic materials have been reported.\cite{CoroIC41}

Recently, the magnetism of pyrimidine bridged transition metal
complexes has been investigated in which pyrimidine
(C$_{4}$H$_{4}$N$_{2}$), in the following abbreviated as PM, plays
the role of an antiferromagnetic coupling
unit.\cite{IshiMCLC233,Munno98,Feyer99,Mohri} Interestingly, the
halide complexes (PM)$_{2}$TX$_{2}$ (T = Co$^{\rm II}$, X = Cl,
Br; T = Fe$^{\rm II}$, X = Cl) possess a chiral 3-dimensional
network of T ions and exhibit weak ferromagnetism below about 5
K.\cite{Naka98,Zusai} The preliminary analysis of magnetization
measurements on these compounds pointed to a very large canting
and an associated large ferromagnetic component of the ordered
moments. It appeared possible that these compounds possess a
chiral magnetically ordered structure.

In order to determine the magnetic structure and to elucidate the
origin of the large canting we performed magnetization
measurements, M\"ossbauer spectroscopy and powder neutron
diffraction on (PM)$_{2}$FeCl$_{2}$. Here, we present a complete
analysis of these data together with electronic structure
calculations. We discuss the implications of our findings for
studies in search of molecule based materials exhibiting chiral
magnetic ordering.

\section{Experimental}

Microcrystalline samples of (PM)$_{2}$FeCl$_{2}$, in the following
abbreviated PFC, were obtained by mixing aqueous solutions of
Fe$^{\rm II}$ chloride (e.g. 10 mmol in 20 ml H$_{2}$O) with the
stoichiometric amount of pyrimidine. The resulting yellow
precipitates were filtered and washed thoroughly with H$_{2}$O.
M\"ossbauer data indicate the presence of a secondary phase, which
we identified as (PM)FeCl$_{2}$. The volume amount of the
secondary phase increases with storage time of the samples. While
a newly made sample contained less than 2\% secondary phase, about
22\% secondary phase have been observed in the same sample stored
for two years under air. This indicates that under such conditions
PFC is unstable against the formation of (PM)FeCl$_{2}$.
Magnetization measurements were carried out using an MPMS Squid
magnetometer (Quantum Design). Some 20 mg of freshly prepared
sample was filled into a gelatine capsule. $^{57}$Fe M\"ossbauer
spectroscopy experiments have been executed in a standard
low-temperature M\"ossbauer set-up at temperatures ranging from 3
K to 300 K (source: $^{57}$Co-in-Rh matrix at room temperature;
half width half maximum (HWHM): 0.130(2) mm/s). The spectra have
been evaluated using the M\"ossbauer fitting program Recoil
\cite{Recoil} in the thin absorber approximation. Above $T_{N}$
the spectra were modeled as Lorentzian lines in the presence of an
electric field gradient. Below $T_{N}$ we employed a method for
the calculation of M\"ossbauer line shapes in the presence of
magnetic dipole and electric quadrupole hyperfine interactions
introduced by Blaes {\it et al}.\cite{Blaes} Neutron powder
diffraction measurements were performed using the instruments E6
and E9 at the Berlin Neutron Scattering Center (BENSC). Neutron
wavelengths of 2.448 \AA~ and 1.7964 \AA, respectively, were used.
The instrument E6 provides a high neutron flux and medium
resolution, is equipped with a 20$^\circ$-multichannel detector,
and covers a range of scattering angles up to about 100$^\circ$.
In contrast, the instrument E9 is a low-flux high-resolution
powder diffractometer with an extended 2$\theta$-range up to
160$^\circ$. Therefore, the former was used for the study of
magnetic Bragg reflections, while the latter was employed for
checking the crystal structure at low temperatures. The
non-deuterated sample was filled into a 7 mm diameter vanadium can
with 40 mm length, resulting in a sample volume of about 1.5
cm$^3$. An absorption correction for cylindrical samples was
applied ($\mu R = 0.94$) to account for the strong incoherent
scattering from hydrogen. During the refinement, no other but
symmetry constraints were used for the atom positional parameters.
Therefore, the validity of the crystal structure model is proven
by the correct geometry of the pyrimidine molecules determined in
the refinement. The Rietveld refinement of the diffraction data
was carried out using the WINPLOTR/FULLPROF
package.\cite{Fullprof}

\section{Theoretical calculations}

The electronic structure calculations were performed with a code
based on a local basis set.\cite{Saun} The unrestricted
Hartree-Fock method and the hybrid functional B3LYP is applied for
solids, and the basis functions are chosen as Gaussian type
orbitals. The iron basis set \cite{Catti} (outermost $d$-exponent
0.4345) is of the size [$5s4p2d$], the chlorine basis set
\cite{Pren} (with one $d$-exponent with value 0.5) of the size
[$5s4p1d$], carbon and nitrogen basis sets \cite{Dove} of the size
[$3s2p1d$], and finally a [$2s1p$] hydrogen basis set \cite{Ditch}
was used. To test the stability of the results, additional tight
basis functions were added for the iron atoms and a diffuse
$sp$-function (exponent 0.12) was added at the nitrogen site to
account for a better description of this negatively charged atom.
The results were, however, found to be essentially stable with
respect to the various basis sets. With these parameters, the
Hartree-Fock or Kohn-Sham equations for a ferro- or
antiferromagnetic (Neel-like) structure are solved
self-consistently, and properties such as charge distributions and
field gradients can be computed. The structural data used in the
calculations is based on the measurements as described in the
following.

\section{Results and discussion}

\subsection{Crystal structure}

The high-resolution powder neutron diffraction data, taken at 10
K, confirm that PFC is isostructural to the Co analogue
\cite{Naka98}, namely tetragonal with the chiral space group
$I4_{1}22$ (or the enantiomorphic $I4_322$) with $a = b =
7.4292(3)$ \AA, $c = 20.364(1)$ \AA. The Fe-N and Fe-Cl distances
2.26 and 2.39 \AA~ are determined, respectively. The crystal
structure is shown in Fig. 1. It consists of a chiral
three-dimensional network of Fe ions coordinated by two Cl and
linked by pyrimidine. Due to the $4_{1}$ (or $4_3$) screw axis,
the local environment of two Fe ions in neighboring layers $z$ and
$z+1/4$ is rotated by 90$^\circ$. We will discuss below that this
specific feature of the crystal structure of PFC can be regarded
as the origin of the large canting observed in the magnetically
ordered state.

In the local FeN$_4$Cl$_2$ geometry all nitrogen atoms of
pyrimidine are equatorially coordinated. It was pointed out
\cite{Mohri} that the electron spins in magnetic $d_{x^2-y^2}$
orbitals are antiferromagnetically correlated through the
pyrimidine molecular orbital(s), i.e., by a superexchange
mechanism,\cite{Good1,Good2,Kana,Gins} when the nitrogen lone pair
and $d_{x^2-y^2}$ orbitals have an appreciable orbital overlap on
both sides.

\begin{figure}

\begin{center}
\epsfig{width=0.43\textwidth,file=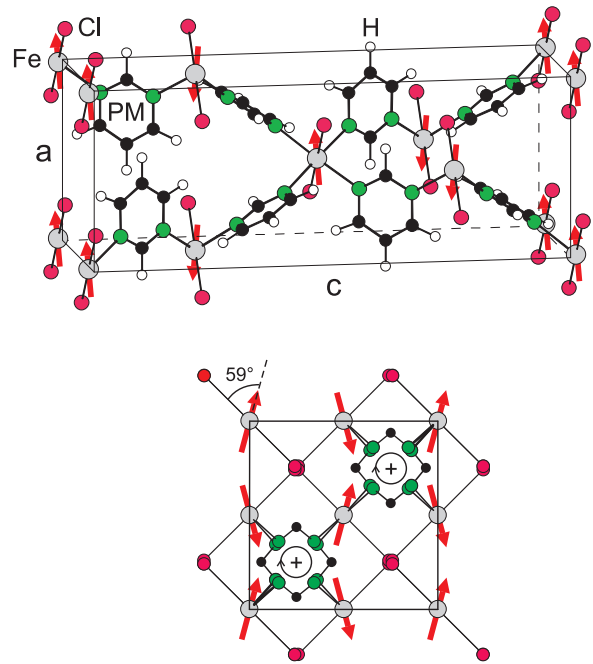}
\end{center}

\caption{Crystal and magnetic structure of PFC as determined from
low-temperature powder neutron diffraction. Lower section: view
along the $c$ axis with the positions of the 90$^\circ$ screw axes
marked. The angle 59$^\circ$ between the magnetic moments and the
local EFG is also marked.} \label{fig:1}
\end{figure}

\section{Results and Discussion}

\subsection{Magnetization}

Figure 2 shows the $M$ vs. $H$ hysteresis curve of PFC measured at
2.0 K. For clarity, the virgin curve is not shown. On increasing
the field from zero, we observe an initial steep increase of the
magnetization in a low field region (a few 100 Oe) to a value much
smaller than the full Fe$^{\rm II}$ moment, and subsequently a
rounded crossover around 500 Oe to a linear field dependence up to
55 kOe (not shown). Such a behavior is typical of a weak
ferromagnet, where the steep increase at low $T$ reflects the
spontaneous magnetization and the linear high-field behavior is
due to strong antiferromagnetic interactions. On sweeping the
field down from 55 kOe, a weak hysteresis develops and a sharp
kink is observed at $H = 0$. This kink allows an accurate
determination of the spontaneous magnetization. We determine a
value of 0.31(1) $\mu_B$ on the powder average. The coercitive
field is small, 150 Oe, and therefore PFC behaves as a soft
magnet. Assuming that the ferromagnetic component in a given
magnetic domain is confined to a specific crystal direction, the
measured powder average of the spontaneous magnetization has to be
multiplied by a factor of three to obtain the ferromagnetic moment
along this axis. This fact is frequently overlooked by other
authors. Thus, the magnetization measurements result in a
ferromagnetic component of 0.93(3) $\mu_B$ per Fe ion in the
canted antiferromagnetic state. We will show in the following that
this value is in full agreement with the combined M\"ossbauer
spectroscopy and neutron diffraction results.

\subsection{M\"ossbauer spectroscopy}

In Fig. 3 we plot the M\"ossbauer-spectra of PFC at temperatures
between 3.75 K and 300 K. At highest temperatures we observe a two
line spectrum, resulting from a quadrupole splitting of $QS =
3.112(3)$ mm/s and an isomer shift of $IS = 0.909(1)$ mm/s
(relative to the $^{57}$Co-in-Rh source). These values are typical
for high spin Fe$^{\rm II}$. Upon lowering the temperature, and
above $T_{N}$, both $QS$ and $IS$ slightly increase, and we find
at 10.4 K values of $QS = 3.271(2)$ mm/s and $IS = 1.037(1)$ mm/s.

\begin{figure}

\begin{center}
\epsfig{width=0.43\textwidth,file=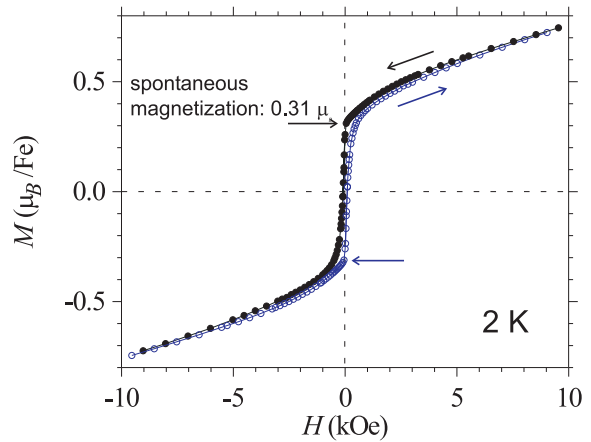}
\end{center}

\caption{Low-field section of the hysteresis-loop of PFC measured
at 2 K. The field was swept between -55 kOe and 55 kOe. For
clarity, the virgin curve is not shown. Horizontal arrows mark
sharp kinks allowing for the determination of the spontaneous
magnetization.} \label{fig:2}
\end{figure}

\begin{figure}
\begin{center}
\epsfig{width=0.40\textwidth,file=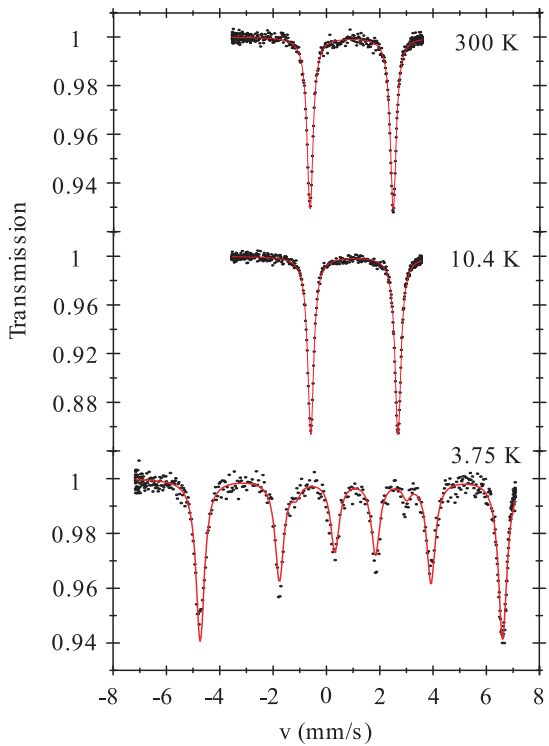}
\end{center}

\caption{M\"ossbauer spectra of PFC measured at various
temperatures. Arrows mark eight absorption lines observed at
3.75~K . The solid lines are fits to the data points (see text).}
\label{fig:3}
\end{figure}

Below T$_{N}$, at 3.75 K, eight separate absorption lines are
observed. These we attribute to the coexistence of an electrical
field gradient (EFG) and a hyperfine magnetic field ($B_{\rm
HF}$). A fit to the data, taking into account the EFG, $B_{\rm
HF}$ and the asymmetry parameter $\eta = (V_{xx}
 - V_{yy})/V_{zz}$ yields values of $IS = 1.035(3)$ mm/s, $\eta = 0.106$, $QS
= 3.28(1)$ mm/s and a hyperfine magnetic field of 30.19(3) T as
best solution. Moreover, from this fit we determine a temperature
independent angle of $B_{\rm HF}$ with the main EFG component
$V_{zz}$ of 59.3(1)$^\circ$ at each Fe site.

\subsection{Results from calculations}

From the calculations, the charges were determined to be +1.9 for
Fe, -1.0 for Cl, -0.8 for N; between 0 and +0.7 for carbon, and
the hydrogen atoms were found to be slightly positive charged (at
most +0.1). The spin at the Fe sites is 1.9 and the magnetic
moment 4.04 $\mu_B$, using a $g$-factor of 2.13.\cite{Zusai} Thus,
there is moment reduction by about 5\% due to covalency effects.
The magnitude of the spin at the Cl, N or C sites is less than 0.1
and thus negligible, and the spin at the hydrogen sites virtually
zero. The electric field gradient at the Fe site has components
with the value $V_{xx} = - 0.8$, $V_{yy} = - 1.3$, and $V_{zz} =
2.1\times 10^{22}$ V/m$^2$. These numbers have an uncertainty of
the order of $0.1\times 10^{22}$ V/m$^2$. The asymmetry parameter
$\eta$ is especially sensitive to this uncertainty and values in
the range from 0.1 to 0.3 are obtained.

The second largest component of the tensor is identical with the
crystallographic $c$-direction, the other two components lie
perpendicular in the tetragonal basal plane and are rotated by
45$^\circ$ with respect to $a$ and $a'$. The largest component is
in the direction of the Cl atoms. We can compare the computed
field gradient with the experimental value by converting according
to the formula $QS = eQV_{zz}c/2E_0$, with $E_0 =14.4$ keV and
using the value of 0.16 barn for the quadrupole moment
$Q$.\cite{Dufek} With these parameters, we obtain a computed $QS =
3.5$ mm/s which is in good agreement with the experimental value.
Finally, it is worthwhile mentioning that the aforementioned
calculated values hardly depend on the magnetic ground state, and
similar values are obtained if a ferromagnetic or an
antiferromagnetic ground state is assumed.

An attempt to extract information about the magnetic exchange
coupling could be made by comparing energies for ferro- and
antiferromagnetic structure. This difference is (per Fe ion)
roughly 0.01 eV at the HF level (with the ferromagnet being
lower), and roughly 0.08 eV at the  B3LYP level (with the
antiferromagnet being lower). However, these energies should be of
the order of the Neel temperature (6.4 K or 0.6 meV). The large
deviation could be due to the fact that a simplified magnetic
order is used in the simulations, neglecting the canting, or that
the description of the electronic correlations is too crude due to
the approximations in the functional. If the HF result was assumed
to be reasonable, then the effect of electronic correlations would
have to be of virtually the same size, but with opposite sign.
Such a subtle effect is very difficult to be computed properly,
and a density functional calculation may be too crude.

\subsection{Magnetic structure}

Figure 4 shows the difference of the powder neutron diffractograms
taken at 1.6 and 7 K (i.e., well below and above $T_{N}$),
revealing the additional Bragg intensities produced by the
long-range magnetic ordering. All magnetic reflections can be
indexed on the basis of the crystallographic unit cell. The rule
$h+k+l$ = even for the observed magnetic Bragg reflections ($hkl$)
indicates that the body centering is conserved. The presence and
strengths of the reflections ($00l$) suggest that the magnetic
moments lie basically perpendicular to the $c$ axis and that
moments in neighboring layers $z$ and $z+1/4$ are basically
oriented antiparallel. Therefore, the antiferromagnetic coupling
through pyrimidine bridges is clearly indicated.

First refinements of the neutron diffraction data resulted in an
ordered moment of roughly 4 $\mu_B$ at 1.6 K. However, the
orientation of the moments within the basal plane as well as the
angle and direction of canting can not be determined from these
data since the only clear signature of the ferromagnetic component
is the weak (112) reflection which is hidden in the noise. We
therefore used the information from the M\"ossbauer spectroscopy
to arrive at the final magnetic structure model.

Assuming the electric field gradient parallel to the Fe-Cl bonds,
as it was determined in the calculations, in the basic
antiferromagnetic structure the moments have to be aligned
parallel to the $a$ axis in order to enclose an identical angle
with the local EFG at each Fe. To increase this angle from
45$^\circ$ to 59.3$^\circ$ - as observed in the M\"ossbauer
experiment - either a canting by 14.3$^\circ$ within the basal
plane (along $a'$) or by 45$^\circ$ along $c$ has to be
introduced. However, only a canting in the basal plane gives a
ferromagnetic component close to 1 $\mu_B$, whereas a canting
along $c$ would require a ferromagnetic component of 2.8 $\mu_B$.
Comparing these values to the results from the magnetization
measurements, the latter model can be clearly ruled out.
Therefore, in the final magnetic structure model, the moments were
confined to the basal plane and the canting angle 14.3$^\circ$ was
fixed. The solid line in Figure 4 is a fit of this final model to
the data. The only free parameter is the magnitude of the ordered
moment. All other parameters, such as lattice constants and
lineshapes, were taken from the fit of the nuclear Bragg pattern
recorded at 7~K (not shown).

Given the large background noise from the hydrogen incoherent
scattering, it gives a good agreement with the data ($R_p =
0.114$, $R_{wp} = 0.137$). The fit yields an ordered moment of
4.0(3) $\mu_B$ - in agreement with the above calculated full
moment on the Fe - and a ferromagnetic component of 1.0(1)
$\mu_B$. The latter value is in good agreement with the
magnetization data.

The temperature dependence of the intensity of the (002) Bragg
reflection is shown in Figure 5 together with the corresponding
data, the square of the hyperfine field $B_{\rm HF}$, from the
M\"ossbauer experiment. It decreases continuously with rising
temperature and vanishes at 6.4(2) K. Assuming that the canting
angle is constant, the Bragg intensity is proportional to the
square of the ordered moment. A fit of the region close to $T_{N}$
with a power law is consistent with a critical exponent of $\beta
= 0.36(1)$ and therefore with 3D Heisenberg behavior.

The canting angle $\alpha = 59.3^\circ - 45^\circ = 14.3^\circ$
observed in the present measurements is extremely large, actually
one of the largest reported to date for any weak ferromagnetic
compound. For example, weak ferromagnetism of Fe with a canting
angle of 16$^\circ$ has been reported for the intermetallic
compound UFe$_{4}$Al$_{8}$. However, the ferromagnetic component
is only 0.3 $\mu_B$ per Fe in this compound and the canting is due
to an interaction between the U 5$f$ and the Fe 3$d$ electrons.
\cite{Paix} Therefore, the physics of this compound is hardly
comparable to that of PFC.

Large canting angles of 2-7$^\circ$ have been also observed in
Gd$_{2}$CuO$_{4}$-type cuprates. For these materials, a
correlation between weak ferromagnetism and the crystal symmetry
has been discussed in detail.\cite{Luo} We believe that such a
correlation is also present in PFC and that the large canting
observed in this compound is a direct result of its chiral lattice
symmetry. Due to this symmetry, the orientation of the local easy
axis varies by 90$^\circ$ between nearest Fe neighbors linked by a
pyrimidine molecule. Therefore, the local anisotropy would favor a
90$^\circ$ alignment between the moments. However, the AFM
interaction favors a 180$^\circ$ alignment. The actual angle
180$^\circ - 2\alpha = 151.4^\circ$ between neighboring moments
therefore can be regarded as the result of the competition of the
local anisotropy and the antiferromagnetic exchange. The final
magnetic structure model suggests that the local easy axis lies
perpendicular to $c$ and within the equatorial plane of the local
coordination octahedron of each Fe. Thus, it coincides with the
axis of the smallest EFG component $V_{xx}$ calculated above. We
may express this anisotropy as an additional term in the
Hamiltonian of the system,

\begin{equation}
\mathbf{H} = \sum\limits_{NN} [J \vec{S}_{1} \cdot \vec{S}_{2} +
D(S_{1x}S_{2y})],
\end{equation}
\noindent where the sum is over all pairs of nearest neighbor
($NN$) spins ($\vec{S}_{1},\vec{S}_{2})$, $\vec{S}_j =
(S_{jx},S_{jy},S_{jz})$, and $D < 0$. The resulting angle $\alpha$
is related to $D$ and $J$ by $D/J = -2\sin 2\alpha$. With the
measured $\alpha$ we get $D = -0.96 J$. If one would assign the
canting to a Dzyaloshinsky-Moriya (DM) type
interaction,\cite{Dzya,Moriya1,Moriya2} the DM vector was aligned
along $c$ as expected for symmetry reasons and the DM term would
read $D(S_{1x}S_{2y} - S_{1y}S_{2x})$. The latter term of (1)
differs from the DM term in leading to a canting towards specific
axes $x$ and $y$ for$S_{1}$ and $S_{2}$, respectively, rather than
a canting within the tetragonal basal plane in general.

It is interesting to note that magnetization data on the Co
analogues to PFC, (PM)$_{2}$CoCl$_{2}$ and
(PM)$_{2}$CoBr$_{2}$,\cite{Naka98} point to even larger
ferromagnetic components and therefore even larger canting angles
than in PFC. Another example of such a large canting may be
3D-[Fe(N$_3$)$_{2}$(4,4'-bpy)], with bpy = 4,4'-bipyridine. A
microcrystalline sample of this compound was recently reported to
exhibit a large spontaneous magnetization of 0.48 $\mu_B$ per Fe
tentatively ascribed to ferromagnetic ordering while a canted
structure was not excluded.\cite{Yuen,Fu} This tetragonal compound
also forms a chiral 3D network structure, space group
$P4_{1}2_{1}2$, which is closely related to that of
(PM)$_{2}$FeCl$_{2}$. The negative Curie-Weiss temperature and the
similarity of the magnetization data of
3D-[Fe(N$_3$)$_{2}$(4,4'-bpy)] with that of the (PM)$_{2}$TX$_{2}$
complexes suggests that also in the former a canted
antiferromagnetic state with a very large canting (of roughly
20$^\circ$) within the tetragonal basal plane is realized.

To our knowledge, only few other molecule based magnets with
chiral crystal structures have been reported. The oxalato (ox)
based compound [Co(2,2'-bpy)$_3$][Co$_2$(ox)$_3$]ClO$_4$
\cite{Herna} exhibits weak ferromagnetism with small canting
($\mu_{\rm fm} = 0.009~\mu_{B}$). The compounds of the series
[Z$^{\rm II}$(2,2'-bpy)$_3$][ClO$_4$][M$^{\rm II}$Cr$^{\rm
III}$(ox)$_3$] \cite{Coro2001} order ferromagnetically. Both
systems crystallize in the cubic chiral space group $P4_132$.
Another example is ferrimagnetic Mn(hfac)$_2$NITPhOMe,
crystallizing in space group $P3_1$. \cite{Cane} The other
compounds discussed in the framework of chiral magnetism are based
on chiral constituents and crystallize in non-centrosymmetric but
achiral space groups, such as $P1$ or $P2_12_12_1$.\cite
{CoroIC41}

\begin{figure}

\begin{center}
\epsfig{width=0.43\textwidth,file=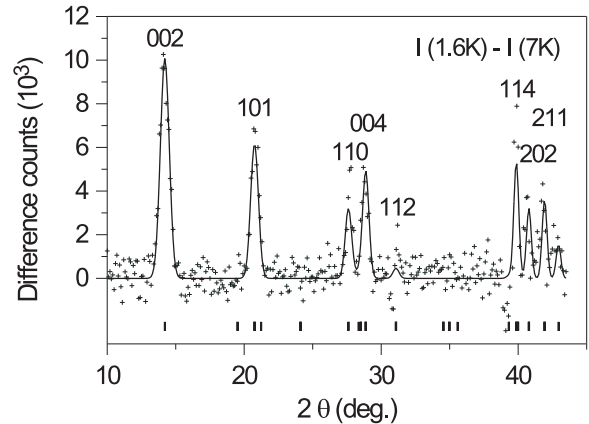}
\end{center}

\caption{Difference of the two neutron powder diffractograms
measured at 1.6 K (below $T_{N}$) and 7.0 K (above $T_{N}$) for
PFC. The wavelength was 2.448 \AA.  The reflections are indexed on
the basis of the crystallographic unit cell.} \label{fig:4}
\end{figure}

\begin{figure}

\begin{center}
\epsfig{width=0.43\textwidth,file=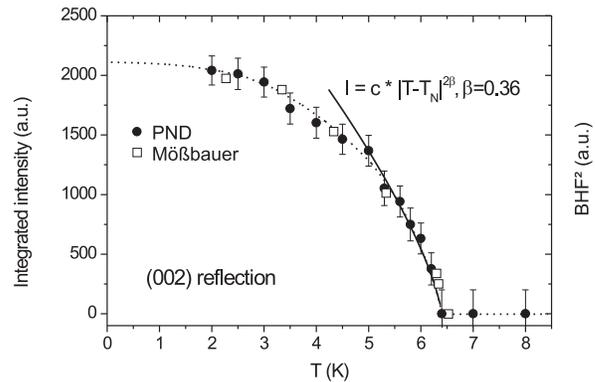}
\end{center}

\caption{Temperature dependence of integrated intensity of the
(002) Bragg reflection, being proportional to the square of the
ordered moment, together with the hyperfine field $B_{\rm HF}$
observed by M\"ossbauer spectroscopy.  The dashed line is a guide
to the eye. The  solid line is a fit to a power law which is
consistent with 3D Heisenberg behavior. } \label{fig:5}

\end{figure}

\section{Conclusion}

We have shown that (PM)$_{2}$FeCl$_{2}$  is a weak ferromagnet
with low coercitive field but a very large ferromagnetic component
of 1 $\mu_B$, corresponding to a canting angle of 14$^\circ$. This
is one of the largest values reported to date. In a given magnetic
domain, the antiferromagnetic component of the moments is confined
to the $a$ axis, while the canting occurs along the perpendicular
direction $a'$. We argued that the large canting in PFC is the
direct result of its chiral lattice symmetry leading to a
competition between the antiferromagnetic exchange interaction and
the local single-ion anisotropy. The appearance of canting rather
than a chiral magnetic structure may be a general feature of the
magnetic ordering in similar structures possessing 90$^\circ$
screw axes. This observation is of interest in view of the
extensive efforts to produce a molecular magnet exhibiting chiral
magnetic ordering. To date, both strategies, crystallizing (i)
achiral constituents in a chiral lattice and (ii) chiral
constituents in non-centrosymmetric but achiral lattices did not
lead to chiral magnetic ordering.

Most of these compounds possess 90$^\circ$ or 180$^\circ$ screw
axes. These lead to a $\pm 90^\circ$ or to no alteration at all,
respectively, of the local anisotropy along the screw axes and
thus do not necessarily support any chirality of the ordered
magnetic structure. Local anisotropy supporting chiral magnetic
ordering therefore appears more likely in chiral
trigonal/hexagonal lattices. We suggest that the search for chiral
magnetic ordering should focus on compounds of that kind.

\begin{acknowledgments}
We thank N. Stuesser and D. Toebbens for experimental support.
This work has been supported by the Deutsche
Forschungsgemeinschaft DFG under contract no. SU229/6-1.
\end{acknowledgments}

\bibliography{fepm2cl2}

\begin{thebibliography}{38}
\expandafter\ifx\csname natexlab\endcsname\relax\def\natexlab#1{#1}\fi
\expandafter\ifx\csname bibnamefont\endcsname\relax
  \def\bibnamefont#1{#1}\fi
\expandafter\ifx\csname bibfnamefont\endcsname\relax
  \def\bibfnamefont#1{#1}\fi
\expandafter\ifx\csname citenamefont\endcsname\relax
  \def\citenamefont#1{#1}\fi
\expandafter\ifx\csname url\endcsname\relax
  \def\url#1{\texttt{#1}}\fi
\expandafter\ifx\csname urlprefix\endcsname\relax\def\urlprefix{URL }\fi
\providecommand{\bibinfo}[2]{#2}
\providecommand{\eprint}[2][]{\url{#2}}

\bibitem[{ICM(2001)}]{ICMM2000}
\emph{\bibinfo{title}{Proc. of ICMM 2000}} (\bibinfo{year}{2001}),
  \bibinfo{note}{in Polyhedron {\bf 20}, pp. 1115--1784}.

\bibitem[{ICM(2003)}]{ICMM2002}
\emph{\bibinfo{title}{Proc. of ICMM 2002}} (\bibinfo{year}{2003}),
  \bibinfo{note}{in Polyhedron, in press}.

\bibitem[{\citenamefont{Kahn}(1993)}]{Kahn}
\bibinfo{author}{\bibfnamefont{O.}~\bibnamefont{Kahn}},
  \emph{\bibinfo{title}{Molecular Magnetism}} (\bibinfo{publisher}{VCH},
  \bibinfo{address}{New York}, \bibinfo{year}{1993}).

\bibitem[{\citenamefont{Miller and Epstein}(1994)}]{Miller}
\bibinfo{author}{\bibfnamefont{J.}~\bibnamefont{Miller}} \bibnamefont{and}
  \bibinfo{author}{\bibfnamefont{A.}~\bibnamefont{Epstein}},
  \bibinfo{journal}{Angew. Chemie} \textbf{\bibinfo{volume}{106}},
  \bibinfo{pages}{399} (\bibinfo{year}{1994}).

\bibitem[{\citenamefont{Gatteschi}(1994)}]{Gatt}
\bibinfo{author}{\bibfnamefont{D.}~\bibnamefont{Gatteschi}},
  \bibinfo{journal}{Adv. Materials} \textbf{\bibinfo{volume}{6}},
  \bibinfo{pages}{635} (\bibinfo{year}{1994}).

\bibitem[{\citenamefont{Coronado et~al.}(2000)\citenamefont{Coronado,
  Gal{\'a}n-Mascar{\'o}s, G{\'o}mez-Garc{\'{\i}}a, and Laukhin}}]{NatureCoro}
\bibinfo{author}{\bibfnamefont{E.}~\bibnamefont{Coronado}},
  \bibinfo{author}{\bibfnamefont{J.~P.} \bibnamefont{Gal{\'a}n-Mascar{\'o}s}},
  \bibinfo{author}{\bibfnamefont{C.~J.} \bibnamefont{G{\'o}mez-Garc{\'{\i}}a}},
  \bibnamefont{and} \bibinfo{author}{\bibfnamefont{V.}~\bibnamefont{Laukhin}},
  \bibinfo{journal}{Nature} \textbf{\bibinfo{volume}{408}},
  \bibinfo{pages}{447} (\bibinfo{year}{2000}).

\bibitem[{\citenamefont{Benard et~al.}(2000)\citenamefont{Benard, Yu,
  Rivi{\`e}re, Cl{\'e}ment, Guilhelm, Tchernatov, and Nakatani}}]{Optic}
\bibinfo{author}{\bibfnamefont{S.}~\bibnamefont{Benard}},
  \bibinfo{author}{\bibfnamefont{P.}~\bibnamefont{Yu}},
  \bibinfo{author}{\bibfnamefont{J.}~\bibnamefont{Rivi{\`e}re}},
  \bibinfo{author}{\bibfnamefont{R.}~\bibnamefont{Cl{\'e}ment}},
  \bibinfo{author}{\bibfnamefont{J.}~\bibnamefont{Guilhelm}},
  \bibinfo{author}{\bibfnamefont{L.}~\bibnamefont{Tchernatov}},
  \bibnamefont{and} \bibinfo{author}{\bibfnamefont{K.}~\bibnamefont{Nakatani}},
  \bibinfo{journal}{J. Am. Chem. Soc.} \textbf{\bibinfo{volume}{122}},
  \bibinfo{pages}{9444} (\bibinfo{year}{2000}).

\bibitem[{\citenamefont{Rikken and Raupach}(1997)}]{Rikken}
\bibinfo{author}{\bibfnamefont{G.~L. J.~A.} \bibnamefont{Rikken}}
  \bibnamefont{and} \bibinfo{author}{\bibfnamefont{E.}~\bibnamefont{Raupach}},
  \bibinfo{journal}{Nature} \textbf{\bibinfo{volume}{390}},
  \bibinfo{pages}{493} (\bibinfo{year}{1997}).

\bibitem[{\citenamefont{Coronado et~al.}(2002)\citenamefont{Coronado,
  G{\'o}mez-Garc{\'{\i}}a, Nuez, Romero, Rusanov, and
  Stoeckli-Evans}}]{CoroIC41}
\bibinfo{author}{\bibfnamefont{E.}~\bibnamefont{Coronado}},
  \bibinfo{author}{\bibfnamefont{C.~J.} \bibnamefont{G{\'o}mez-Garc{\'{\i}}a}},
  \bibinfo{author}{\bibfnamefont{A.}~\bibnamefont{Nuez}},
  \bibinfo{author}{\bibfnamefont{F.~M.} \bibnamefont{Romero}},
  \bibinfo{author}{\bibfnamefont{E.}~\bibnamefont{Rusanov}}, \bibnamefont{and}
  \bibinfo{author}{\bibfnamefont{H.}~\bibnamefont{Stoeckli-Evans}},
  \bibinfo{journal}{Inorg. Chem.} \textbf{\bibinfo{volume}{41}},
  \bibinfo{pages}{4615} (\bibinfo{year}{2002}), \bibinfo{note}{and references
  therein}.

\bibitem[{\citenamefont{Ishida et~al.}(1993)\citenamefont{Ishida, Mitsubori,
  Nogami, and Iwamura}}]{IshiMCLC233}
\bibinfo{author}{\bibfnamefont{T.}~\bibnamefont{Ishida}},
  \bibinfo{author}{\bibfnamefont{S.-I.} \bibnamefont{Mitsubori}},
  \bibinfo{author}{\bibfnamefont{T.}~\bibnamefont{Nogami}}, \bibnamefont{and}
  \bibinfo{author}{\bibfnamefont{H.}~\bibnamefont{Iwamura}},
  \bibinfo{journal}{Mol. Cryst. Liq. Cryst.} \textbf{\bibinfo{volume}{233}},
  \bibinfo{pages}{345} (\bibinfo{year}{1993}).

\bibitem[{\citenamefont{Munno et~al.}(1998)\citenamefont{Munno, Poerio, Julve,
  Lloret, and Viau}}]{Munno98}
\bibinfo{author}{\bibfnamefont{G.~D.} \bibnamefont{Munno}},
  \bibinfo{author}{\bibfnamefont{T.}~\bibnamefont{Poerio}},
  \bibinfo{author}{\bibfnamefont{M.}~\bibnamefont{Julve}},
  \bibinfo{author}{\bibfnamefont{F.}~\bibnamefont{Lloret}}, \bibnamefont{and}
  \bibinfo{author}{\bibfnamefont{G.}~\bibnamefont{Viau}}, \bibinfo{journal}{New
  J. Chem.} p. \bibinfo{pages}{299} (\bibinfo{year}{1998}).

\bibitem[{\citenamefont{Feyerherm et~al.}(2000)\citenamefont{Feyerherm, Abens,
  G{\"u}nther, Ishida, Mei{\ss}ner, Meschke, and Nogami}}]{Feyer99}
\bibinfo{author}{\bibfnamefont{R.}~\bibnamefont{Feyerherm}},
  \bibinfo{author}{\bibfnamefont{S.}~\bibnamefont{Abens}},
  \bibinfo{author}{\bibfnamefont{D.}~\bibnamefont{G{\"u}nther}},
  \bibinfo{author}{\bibfnamefont{T.}~\bibnamefont{Ishida}},
  \bibinfo{author}{\bibfnamefont{M.}~\bibnamefont{Mei{\ss}ner}},
  \bibinfo{author}{\bibfnamefont{M.}~\bibnamefont{Meschke}}, \bibnamefont{and}
  \bibinfo{author}{\bibfnamefont{T.}~\bibnamefont{Nogami}},
  \bibinfo{journal}{J. Phys.: Condens. Matter.} \textbf{\bibinfo{volume}{39}},
  \bibinfo{pages}{8495} (\bibinfo{year}{2000}).

\bibitem[{\citenamefont{Mohri et~al.}(1999)\citenamefont{Mohri, Yoshizawa,
  Yamabe, Ishida, and Nogami}}]{Mohri}
\bibinfo{author}{\bibfnamefont{F.}~\bibnamefont{Mohri}},
  \bibinfo{author}{\bibfnamefont{K.}~\bibnamefont{Yoshizawa}},
  \bibinfo{author}{\bibfnamefont{T.}~\bibnamefont{Yamabe}},
  \bibinfo{author}{\bibfnamefont{T.}~\bibnamefont{Ishida}}, \bibnamefont{and}
  \bibinfo{author}{\bibfnamefont{T.}~\bibnamefont{Nogami}},
  \bibinfo{journal}{Mol. Engineer.} \textbf{\bibinfo{volume}{8}},
  \bibinfo{pages}{357} (\bibinfo{year}{1999}).

\bibitem[{\citenamefont{Nakayama et~al.}(1998)\citenamefont{Nakayama, Ishida,
  Takayama, Hashizume, Yasui, Iwasaki, and Nogami}}]{Naka98}
\bibinfo{author}{\bibfnamefont{K.}~\bibnamefont{Nakayama}},
  \bibinfo{author}{\bibfnamefont{T.}~\bibnamefont{Ishida}},
  \bibinfo{author}{\bibfnamefont{R.}~\bibnamefont{Takayama}},
  \bibinfo{author}{\bibfnamefont{D.}~\bibnamefont{Hashizume}},
  \bibinfo{author}{\bibfnamefont{M.}~\bibnamefont{Yasui}},
  \bibinfo{author}{\bibfnamefont{F.}~\bibnamefont{Iwasaki}}, \bibnamefont{and}
  \bibinfo{author}{\bibfnamefont{T.}~\bibnamefont{Nogami}},
  \bibinfo{journal}{Chem. Lett.} p. \bibinfo{pages}{497}
  (\bibinfo{year}{1998}).

\bibitem[{\citenamefont{Zusai et~al.}(2000)\citenamefont{Zusai, Kusaka, Ishida,
  Feyerherm, Steiner, and Nogami}}]{Zusai}
\bibinfo{author}{\bibfnamefont{K.}~\bibnamefont{Zusai}},
  \bibinfo{author}{\bibfnamefont{T.}~\bibnamefont{Kusaka}},
  \bibinfo{author}{\bibfnamefont{T.}~\bibnamefont{Ishida}},
  \bibinfo{author}{\bibfnamefont{R.}~\bibnamefont{Feyerherm}},
  \bibinfo{author}{\bibfnamefont{M.}~\bibnamefont{Steiner}}, \bibnamefont{and}
  \bibinfo{author}{\bibfnamefont{T.}~\bibnamefont{Nogami}},
  \bibinfo{journal}{Mol. Cryst. Liq. Cryst.} \textbf{\bibinfo{volume}{343}},
  \bibinfo{pages}{127} (\bibinfo{year}{2000}).

\bibitem[{\citenamefont{Lagarec}(1998)}]{Recoil}
\bibinfo{author}{\bibfnamefont{K.}~\bibnamefont{Lagarec}},
  \bibinfo{journal}{Department of Physics University of Ottawa}
  \textbf{\bibinfo{volume}{Version 1.02}} (\bibinfo{year}{1998}).

\bibitem[{\citenamefont{Blaes et~al.}(1985)\citenamefont{Blaes, Fischer, and
  Gonser}}]{Blaes}
\bibinfo{author}{\bibfnamefont{N.}~\bibnamefont{Blaes}},
  \bibinfo{author}{\bibfnamefont{H.}~\bibnamefont{Fischer}}, \bibnamefont{and}
  \bibinfo{author}{\bibfnamefont{U.}~\bibnamefont{Gonser}},
  \bibinfo{journal}{Nucl. Instr. and Meth. in Phys. Res. B}
  \textbf{\bibinfo{volume}{9}}, \bibinfo{pages}{201} (\bibinfo{year}{1985}).

\bibitem[{\citenamefont{Roisnel and
  Rodr{\'{\i}}guez-Carvajal}(2001)}]{Fullprof}
\bibinfo{author}{\bibfnamefont{T.}~\bibnamefont{Roisnel}} \bibnamefont{and}
  \bibinfo{author}{\bibfnamefont{J.}~\bibnamefont{Rodr{\'{\i}}guez-Carvajal}},
  \bibinfo{journal}{Materials Science Forum}
  \textbf{\bibinfo{volume}{378-381}}, \bibinfo{pages}{118}
  (\bibinfo{year}{2001}).

\bibitem[{\citenamefont{Saunders et~al.}(1998)\citenamefont{Saunders, Dovesi,
  Roetti, Caus{\'a}, Harrison, Orlando, and Zicovich-Wilson}}]{Saun}
\bibinfo{author}{\bibfnamefont{V.~R.} \bibnamefont{Saunders}},
  \bibinfo{author}{\bibfnamefont{R.}~\bibnamefont{Dovesi}},
  \bibinfo{author}{\bibfnamefont{C.}~\bibnamefont{Roetti}},
  \bibinfo{author}{\bibfnamefont{M.}~\bibnamefont{Caus{\'a}}},
  \bibinfo{author}{\bibfnamefont{N.~M.} \bibnamefont{Harrison}},
  \bibinfo{author}{\bibfnamefont{R.}~\bibnamefont{Orlando}}, \bibnamefont{and}
  \bibinfo{author}{\bibfnamefont{C.~M.} \bibnamefont{Zicovich-Wilson}},
  \emph{\bibinfo{title}{CRYSTAL98 User's Manual}},
  \bibinfo{address}{Theoretical Chemistry Group, University of Torino}
  (\bibinfo{year}{1998}).

\bibitem[{\citenamefont{Catti et~al.}(1995)\citenamefont{Catti, G.Valerio, and
  Dovesi}}]{Catti}
\bibinfo{author}{\bibfnamefont{M.}~\bibnamefont{Catti}},
  \bibinfo{author}{\bibnamefont{G.Valerio}}, \bibnamefont{and}
  \bibinfo{author}{\bibfnamefont{R.}~\bibnamefont{Dovesi}},
  \bibinfo{journal}{Phys. Rev. B} \textbf{\bibinfo{volume}{51}},
  \bibinfo{pages}{7441} (\bibinfo{year}{1995}).

\bibitem[{\citenamefont{Prencipe et~al.}(1995)\citenamefont{Prencipe, Zupan,
  Dovesi, Apr{\'a}, and Saunders}}]{Pren}
\bibinfo{author}{\bibfnamefont{M.}~\bibnamefont{Prencipe}},
  \bibinfo{author}{\bibfnamefont{A.}~\bibnamefont{Zupan}},
  \bibinfo{author}{\bibfnamefont{R.}~\bibnamefont{Dovesi}},
  \bibinfo{author}{\bibfnamefont{E.}~\bibnamefont{Apr{\'a}}}, \bibnamefont{and}
  \bibinfo{author}{\bibfnamefont{V.~R.} \bibnamefont{Saunders}},
  \bibinfo{journal}{Phys. Rev. B} \textbf{\bibinfo{volume}{51}},
  \bibinfo{pages}{3391} (\bibinfo{year}{1995}).

\bibitem[{\citenamefont{Dovesi et~al.}(1990)\citenamefont{Dovesi, Caus{\'a},
  Orlando, Roetti, and Saunders}}]{Dove}
\bibinfo{author}{\bibfnamefont{R.}~\bibnamefont{Dovesi}},
  \bibinfo{author}{\bibfnamefont{M.}~\bibnamefont{Caus{\'a}}},
  \bibinfo{author}{\bibfnamefont{R.}~\bibnamefont{Orlando}},
  \bibinfo{author}{\bibfnamefont{C.}~\bibnamefont{Roetti}}, \bibnamefont{and}
  \bibinfo{author}{\bibfnamefont{V.~R.} \bibnamefont{Saunders}},
  \bibinfo{journal}{J. Chem. Phys.} \textbf{\bibinfo{volume}{92}},
  \bibinfo{pages}{7402} (\bibinfo{year}{1990}).

\bibitem[{\citenamefont{Ditchfield et~al.}(1971)\citenamefont{Ditchfield,
  Hehre, and Pople}}]{Ditch}
\bibinfo{author}{\bibfnamefont{R.}~\bibnamefont{Ditchfield}},
  \bibinfo{author}{\bibfnamefont{W.~J.} \bibnamefont{Hehre}}, \bibnamefont{and}
  \bibinfo{author}{\bibfnamefont{J.~A.} \bibnamefont{Pople}},
  \bibinfo{journal}{J. Chem. Phys.} \textbf{\bibinfo{volume}{54}},
  \bibinfo{pages}{724} (\bibinfo{year}{1971}).

\bibitem[{\citenamefont{Kanamori}(1959)}]{Kana}
\bibinfo{author}{\bibfnamefont{J.}~\bibnamefont{Kanamori}},
  \bibinfo{journal}{J. Phys. Chem. Solids} \textbf{\bibinfo{volume}{10}},
  \bibinfo{pages}{87} (\bibinfo{year}{1959}).

\bibitem[{\citenamefont{Ginsberg}(1971)}]{Gins}
\bibinfo{author}{\bibfnamefont{A.~P.} \bibnamefont{Ginsberg}},
  \bibinfo{journal}{Inorg. Chim. Acta Rev.} \textbf{\bibinfo{volume}{5}},
  \bibinfo{pages}{45} (\bibinfo{year}{1971}).

\bibitem[{\citenamefont{Goodenough}(1955)}]{Good1}
\bibinfo{author}{\bibfnamefont{J.~B.} \bibnamefont{Goodenough}},
  \bibinfo{journal}{Phys. Rev.} \textbf{\bibinfo{volume}{100}},
  \bibinfo{pages}{564} (\bibinfo{year}{1955}).

\bibitem[{\citenamefont{Goodenough}(1958)}]{Good2}
\bibinfo{author}{\bibfnamefont{J.~B.} \bibnamefont{Goodenough}},
  \bibinfo{journal}{J. Phys. Chem. Solids} \textbf{\bibinfo{volume}{6}},
  \bibinfo{pages}{287} (\bibinfo{year}{1958}).

\bibitem[{\citenamefont{Dufek et~al.}(1995)\citenamefont{Dufek, Blaha, and
  Schwarz}}]{Dufek}
\bibinfo{author}{\bibfnamefont{P.}~\bibnamefont{Dufek}},
  \bibinfo{author}{\bibfnamefont{P.}~\bibnamefont{Blaha}}, \bibnamefont{and}
  \bibinfo{author}{\bibfnamefont{K.}~\bibnamefont{Schwarz}},
  \bibinfo{journal}{Phys. Rev. Lett.} \textbf{\bibinfo{volume}{75}},
  \bibinfo{pages}{3545} (\bibinfo{year}{1995}).

\bibitem[{\citenamefont{Paix{\~a}o et~al.}(1997)\citenamefont{Paix{\~a}o,
  Lebech, Gon{\c{c}}alves, Brown, Lander, Burlet, Delapalme, and
  Spirlet}}]{Paix}
\bibinfo{author}{\bibfnamefont{J.~A.} \bibnamefont{Paix{\~a}o}},
  \bibinfo{author}{\bibfnamefont{B.}~\bibnamefont{Lebech}},
  \bibinfo{author}{\bibfnamefont{A.~P.} \bibnamefont{Gon{\c{c}}alves}},
  \bibinfo{author}{\bibfnamefont{P.~J.} \bibnamefont{Brown}},
  \bibinfo{author}{\bibfnamefont{G.~H.} \bibnamefont{Lander}},
  \bibinfo{author}{\bibfnamefont{P.}~\bibnamefont{Burlet}},
  \bibinfo{author}{\bibfnamefont{A.}~\bibnamefont{Delapalme}},
  \bibnamefont{and} \bibinfo{author}{\bibfnamefont{J.~C.}
  \bibnamefont{Spirlet}}, \bibinfo{journal}{Phys. Rev. B}
  \textbf{\bibinfo{volume}{55}}, \bibinfo{pages}{14370} (\bibinfo{year}{1997}).

\bibitem[{\citenamefont{Luo et~al.}(1999)\citenamefont{Luo, Hsu, Lin, Chi, Lee,
  and Ku}}]{Luo}
\bibinfo{author}{\bibfnamefont{H.~M.} \bibnamefont{Luo}},
  \bibinfo{author}{\bibfnamefont{Y.~Y.} \bibnamefont{Hsu}},
  \bibinfo{author}{\bibfnamefont{B.~N.} \bibnamefont{Lin}},
  \bibinfo{author}{\bibfnamefont{P.}~\bibnamefont{Chi}},
  \bibinfo{author}{\bibfnamefont{T.~J.} \bibnamefont{Lee}}, \bibnamefont{and}
  \bibinfo{author}{\bibfnamefont{H.~C.} \bibnamefont{Ku}},
  \bibinfo{journal}{Phys. Rev. B} \textbf{\bibinfo{volume}{60}},
  \bibinfo{pages}{13119} (\bibinfo{year}{1999}).

\bibitem[{\citenamefont{Dzyaloshinsky}(1958)}]{Dzya}
\bibinfo{author}{\bibfnamefont{I.}~\bibnamefont{Dzyaloshinsky}},
  \bibinfo{journal}{J. Phys. Chem. Solids} \textbf{\bibinfo{volume}{4}},
  \bibinfo{pages}{241} (\bibinfo{year}{1958}).

\bibitem[{\citenamefont{Moriya}(1960{\natexlab{a}})}]{Moriya1}
\bibinfo{author}{\bibfnamefont{T.}~\bibnamefont{Moriya}},
  \bibinfo{journal}{Phys. Rev.} \textbf{\bibinfo{volume}{117}},
  \bibinfo{pages}{635} (\bibinfo{year}{1960}{\natexlab{a}}).

\bibitem[{\citenamefont{Moriya}(1960{\natexlab{b}})}]{Moriya2}
\bibinfo{author}{\bibfnamefont{T.}~\bibnamefont{Moriya}},
  \bibinfo{journal}{Phys. Rev.} \textbf{\bibinfo{volume}{120}},
  \bibinfo{pages}{91} (\bibinfo{year}{1960}{\natexlab{b}}).

\bibitem[{\citenamefont{Yuen et~al.}(2002)\citenamefont{Yuen, Lin, Fu, and
  Li}}]{Yuen}
\bibinfo{author}{\bibfnamefont{T.}~\bibnamefont{Yuen}},
  \bibinfo{author}{\bibfnamefont{C.~L.} \bibnamefont{Lin}},
  \bibinfo{author}{\bibfnamefont{A.}~\bibnamefont{Fu}}, \bibnamefont{and}
  \bibinfo{author}{\bibfnamefont{J.}~\bibnamefont{Li}}, \bibinfo{journal}{J.
  Appl. Phys.} \textbf{\bibinfo{volume}{91}}, \bibinfo{pages}{7385}
  (\bibinfo{year}{2002}).

\bibitem[{\citenamefont{Fu et~al.}(2002)\citenamefont{Fu, Huang, Li, Yuen, and
  Lin}}]{Fu}
\bibinfo{author}{\bibfnamefont{A.}~\bibnamefont{Fu}},
  \bibinfo{author}{\bibfnamefont{X.}~\bibnamefont{Huang}},
  \bibinfo{author}{\bibfnamefont{J.}~\bibnamefont{Li}},
  \bibinfo{author}{\bibfnamefont{T.}~\bibnamefont{Yuen}}, \bibnamefont{and}
  \bibinfo{author}{\bibfnamefont{C.~L.} \bibnamefont{Lin}},
  \bibinfo{journal}{Chem. Eur. J.} \textbf{\bibinfo{volume}{8}},
  \bibinfo{pages}{2239} (\bibinfo{year}{2002}).

\bibitem[{\citenamefont{Hern{\'a}ndez-Molina
  et~al.}(1998)\citenamefont{Hern{\'a}ndez-Molina, Lloret, Ruiz-P{\'e}rez, and
  Julve}}]{Herna}
\bibinfo{author}{\bibfnamefont{M.}~\bibnamefont{Hern{\'a}ndez-Molina}},
  \bibinfo{author}{\bibfnamefont{F.}~\bibnamefont{Lloret}},
  \bibinfo{author}{\bibfnamefont{C.}~\bibnamefont{Ruiz-P{\'e}rez}},
  \bibnamefont{and} \bibinfo{author}{\bibfnamefont{M.}~\bibnamefont{Julve}},
  \bibinfo{journal}{Inorg. Chem.} \textbf{\bibinfo{volume}{37}},
  \bibinfo{pages}{4131} (\bibinfo{year}{1998}).

\bibitem[{\citenamefont{Coronado et~al.}(2001)\citenamefont{Coronado,
  Gal{\'a}n-Mascar{\'o}s, G{\'o}mez-Garc{\'{\i}}a, and
  Mart{\'{\i}}nez-Agudo}}]{Coro2001}
\bibinfo{author}{\bibfnamefont{E.}~\bibnamefont{Coronado}},
  \bibinfo{author}{\bibfnamefont{J.~P.} \bibnamefont{Gal{\'a}n-Mascar{\'o}s}},
  \bibinfo{author}{\bibfnamefont{C.~J.} \bibnamefont{G{\'o}mez-Garc{\'{\i}}a}},
  \bibnamefont{and} \bibinfo{author}{\bibfnamefont{J.~M.}
  \bibnamefont{Mart{\'{\i}}nez-Agudo}}, \bibinfo{journal}{Inorg. Chem.}
  \textbf{\bibinfo{volume}{40}}, \bibinfo{pages}{113} (\bibinfo{year}{2001}).

\bibitem[{\citenamefont{Caneschi et~al.}(1991)\citenamefont{Caneschi,
  Gatteschi, Rey, and Sessoli}}]{Cane}
\bibinfo{author}{\bibfnamefont{A.}~\bibnamefont{Caneschi}},
  \bibinfo{author}{\bibfnamefont{D.}~\bibnamefont{Gatteschi}},
  \bibinfo{author}{\bibfnamefont{P.}~\bibnamefont{Rey}}, \bibnamefont{and}
  \bibinfo{author}{\bibfnamefont{R.}~\bibnamefont{Sessoli}},
  \bibinfo{journal}{Inorg. Chem.} \textbf{\bibinfo{volume}{30}},
  \bibinfo{pages}{3936} (\bibinfo{year}{1991}).

\end{thebibliography}

\end{document}